\documentstyle[prl,aps,psfig,epsf,floats,twocolumn]{revtex}

\newcommand{\avg}[1]{\langle{#1}\rangle}

\newcommand{\beq}{\begin{equation}}
\newcommand{\eeq}{\end{equation}}
\newcommand{\beqas}{\begin{eqnarray*}}
\newcommand{\eeqas}{\end{eqnarray*}}

\begin{document}
\draft
\twocolumn[\hsize\textwidth\columnwidth\hsize\csname 
@twocolumnfalse\endcsname

\title{Generalized Dielectric Breakdown Model}
\author{R. Cafiero $^{1,2}$,  A. Gabrielli  $^{2,3}$,
M. Marsili $^{4}$, M. A. Mu\~noz $^{5,2}$ and L. Pietronero $^{2}$.}
\address{ $^{1}$ Max-Planck Institute for Physics of Complex Systems,
N\"othnitzer Strasse 38, 01187 Dresden (Germany)}
\address{$^{2}$ Dipartimento di Fisica e unit\`a INFM, Universit\`a degli 
Studi ``La Sapienza'' , P.le A. Moro 2, 00185 Roma, Italy}
\address{$^{3}$Dipartimento di Fisica Universit\`a degli Studi ``Tor Vergata'',
v.le della Ricerca Scientifica 1, 00133 Roma Italy}
\address{ $^{4}$
 International School For Advanced Studies (SISSA), and unit\`a INFM,
 V. Beirut 2-4, 
34014 Trieste, Italy}
\address{ $^{5}$ The Abdus Salam International
 Centre for Theoretical Physics (ICTP)
P.O. Box 586, 34100 Trieste, Italy}

\maketitle
\begin{abstract}
We propose a generalized version of the Dielectric
Breakdown Model (DBM) for generic breakdown processes. It
interpolates between the 
standard DBM and its 
analog with quenched disorder,
as a temperature like parameter is varied. The physics
of other well known fractal growth phenomena as
Invasion Percolation and the Eden model are also
recovered for some particular parameter values.
The competition between different growing mechanisms
leads to new non-trivial effects and allows us to better 
describe real growth phenomena.
 Detailed 
numerical and theoretical analysis are performed to 
study the interplay between the elementary 
mechanisms. 
In particular, we
observe a continuously changing fractal dimension 
as temperature is varied, and report an
evidence of a  novel phase transition at zero temperature 
in absence of an external driving field; the temperature
acts as a relevant parameter for the ``self-organized''
invasion percolation fixed point. This permits us to obtain
new insight into the connections between self-organization and
standard phase transitions.    
\end{abstract}

\pacs{61.43.-j,61.43.Hv,02.50.+s.}

\narrowtext
\vskip2pc]

Fractal growth phenomena, such as viscous fingering,
electric discharges in dielectrics, fracture propagation and
fluid flow in porous media, have attracted much attention
in recent years. In the study of these phenomena many models -- the
Diffusion Limited Aggregation (DLA) \cite{dla}, Dielectric Breakdown Model
(DBM) \cite{dbm}, and Invasion Percolation (IP)\cite{wilk} 
 to quote only but a few --
have been introduced as a first step towards the understanding of
the dynamic emergence of fractal structures in nature.
These models have been quite successful in reproducing
the essence of the above complex phenomena by capturing 
some key ingredients.
 Their simplicity has also allowed for analytical
studies of the scale invariant properties of corresponding
 structures \cite{rtslung,fst2}.

  To proceed further in this field there are two possible
directions: One is to identify possible interrelations between these 
models and understand their eventual universality or fundamental 
differences. For example, 
these irreversible processes have a much lower degree of
universality than their counterparts in equilibrium statistical mechanics,
 but still the basic ``dynamical screening'' mechanisms
leading to fractal growth (as opposed to the growth of 
compact clusters) can be categorized in some generic classes,
among them that arising from the external physical field
 (e.g.
a Laplacian field in the DBM), or extremal dynamics
and memory effects \cite{rtslung,mem}.
The second is to make these models more realistic or closer to 
real phenomena.  For example, 
the interplay between these different
mechanisms has been hardly studied
so far. For instance, it has been recently found that the
combination of Laplacian screening and extremal dynamics,
 results in a much stronger screening effect
than that of each single mechanism, leading to 
fractal dimensions lower than those characterizing the 
two different elementary mechanisms
\cite{qdbmlet}. 

 Here we present a step forward in both of the previous
 directions. 
 On one hand, we introduce a generalization of the available
models in a broader 
class which permits us the identification 
of relevant parameters 
 in more realistic growth phenomena.
 \begin{figure}
\centerline{\psfig{figure=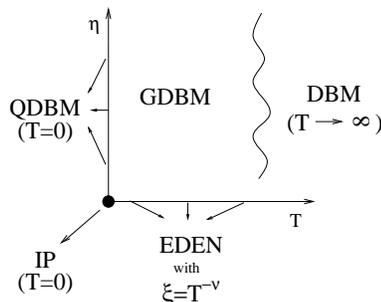,height=4cm,angle=-90}}
\caption{Schematic diagram showing the limiting cases of
our model as a function of temperature and
the parameter $\eta$.
 }
\label{fig0}
\end{figure}
 On the other hand this
broader class is theoretically important as it is able to cast in a 
coherent framework apparently unrelated physical situations.
This gives a new perspective that permits us to clarify the roles
of the single physical mechanisms in competition, namely
external driving fields,
quenched disorder,
and thermal noise.
The relative strengths of these effects
are modulated in our model by 
three parameters, $\eta$, $a$ and $T$ respectively.
For some particular values of these parameters different well
known
growth models are recovered (see figure \ref{fig0}). 
For $T=0$ we have QDBM, in the infinite
$T$ limit the model reduces to the DBM; at
$\eta=0$ and zero temperature we have IP, while for generic
temperatures we have compact (EDEN \cite{EDEN}) growth with a temperature
dependent correlation length. 

Let us now present the model, 
 that we name as
Generalized Dielectric Breakdown Model (GDBM). 
The medium in which
a breakdown process propagates is discretized 
as a regular (say hypercubic) lattice.
The relevant degrees of freedom  are placed on the bonds
connecting lattice sites.
 Each of these bonds can either
be broken or not. Unbroken bonds are characterized by a
given variable, $s_i$, that can take different values
(for instance, this can represent in a sketchy way the
elongation of a spring at $i$).
 The simplest situation
one can think of is that of $s_i$ being
 a spin-like variable $s_i=\pm 1$.
We consider a cylindrical geometry
with lateral periodic boundary conditions.
This geometry, in fact, avoids all complications due to the
discretized nature of the lattice, (like the anisotropy
of growth along the diagonals of the lattice, 
relevant in geometries with radial symmetry \cite{fst2}).
We are interested in studying the process by which the bonds
in the interface break down successively,
 and in how the cluster of broken bonds evolves.
 
  The dynamics proceeds as follows. First of all we assume that 
the breakdown process is quasi-static, i.e. only one bond is broken 
at each time step, and it 
 is in the interface of previously unbroken bonds, this 
is, the cluster of broken bonds is connected.
As initial condition we take as broken all the bonds in the
lower row of the cylinder.
 Secondly,
     there are two distinct processes going
on during the breakdown, the time  scales of which are widely
separated. The first, slow process, is the dynamics of breakdown
itself. The second, fast process, is the relaxation of the field
configuration and of variables $s_i$
after a breakdown event.
 In between two consecutives 
breakings the bond variables relax to their associated equilibrium 
state, whose associated energy or Hamiltonian will be defined
later.
 The separation of time-scales
is a common ingredient in many models for fractal growth
and self-organized criticality \cite{ddrg}, and is physically a quite
reasonable assumption.

We now define the fast dynamics in between two consecutive
breakdowns.
Every bond $i$ is subject to
a stress Laplacian field $E_i$. As in the DBM this (electric) field
is given by the
solution of the Laplace equation with the appropriate
boundary conditions on the growing fracture \cite{dbm,fst2}, namely
the potential is $1$ at the upper border of the cylinder 
(upper electrode) and $0$
on the lower one and on the 
broken bonds.
This field acts over the bonds 
modulated by a parameter $\eta$ \cite{dbm},
 $E_i^{\eta}$, for
$\eta=0$ any dependency on the external field is canceled.
         
 The second physical ingredient is quenched disorder.
At each bond we define a random resistance
 $x_i$, distributed
as $\pi(x)=ax^{a-1}$,
 $x\in[0,1]$ and $a\in[0,\infty)$. This
 gives an idea of the bond tolerance to applied stresses;
the larger $x_i$ the larger the resistance.
  Other probability distributions 
 can be used without affecting the
properties of the model.
 We now define an effective local field accounting for
the electric field and the disorder
$h_i=E_i^\eta/x_i$ 
which acts on the local 
degree of freedom $s_i$        
generating an interaction energy given by
\begin{equation} H_i=h_i s_i.
\label{hamileff}
\end{equation}
There
are looser bonds, with $x_i$ large, which can accommodate
a larger stress $E_i^\eta$ induced by the Laplacian
field, and tighter, or more fragile, bonds, with $x_i$ 
small.

In general one could also include nearest neighbor
interactions, so that $H_i=h_i s_i+\sum_{j} J s_i 
s_j$ to account for material rigidity.
 However, for the situations we are interested in, the 
local stress fields are much larger than the
coupling $J$ and $J$ can be neglected.

The equilibrium statistical
mechanics of each bond variable among
two successive breakdown processes is easily determined by
introducing the temperature $T$.
The partition function 
factorizes on the bonds: $Z(\{h_i\},T)=\prod_i z(h_i,T)$
and 
\[z(h_i,T)=\sum_{s_i=\pm 1}e^{-h_i s_i/T}=2 \cosh 
(h_i/T).\]
We can now calculate, as a function of $h_i$ and $T$,
all averages. In particular 
\[\avg{s_i}=-T\frac{\partial}{\partial h_i}\ln 
z(h_i,T)=\tanh (h_i/T)\]
\[\avg{\delta s_i^2}=1-\tanh^2 (h_i/T)\]
are of interest.
It is clear that $\avg{s_i}$ 
is a measure
of the stress exerted by the external field 
on the bond $i$ and $\avg{\delta 
s_i^2}$ measures
the strength of thermal fluctuations.

We consider the breaking probability 
$P_i$ of a bond as an increasing function 
of the stress, and as a decreasing function of
 the amplitude of its thermal 
fluctuations; 
 it is indeed when the local degree of freedom 
is thermally frozen in a stressed state, i.e. when it is very rigid, 
that breakdown is more likely to occur. On the other hand when the
variance is large, the spin can better absorb stress and is more
flexible.

Guided by these considerations, 
the simplest dimensionless expression for the
breaking probability $P_i$ is
\begin{equation}
P_i\propto\frac{\avg{s_i}}{\sqrt{\avg{\delta s_i^2}}}=\sinh 
(h_i/T).
\label{pi}
\end{equation}
At each time step all the $P_i$ are calculated, and with 
the corresponding probabilities one bond is broken.
After that, the boundary conditions are automatically
modified and the field $E$ 
updated at each point. Using the hypothesis of time scale
separation,  
$z(h_i,T)$ and $P_i$ are automatically updated, a new 
bond breaks down
in the interface, and the process is iterated.

The behavior of the model in some limiting cases can be 
analytically sorted out.

(i) For $T\gg 1$ one can expand 
Eq. (\ref{pi}) around $h_i=0$,
 and find $P_i \propto E_i^\eta/x_i$. In this limit 
the disorder does not play
a relevant role and one obtains back the DBM (this same fact was 
already observed in \cite{qdbm}). 

(ii) For $T \ll 1$, on the other hand, the argument of the
$\sinh$ is very large and essentially only the bond with
the largest $h_i$ has a finite probability to break
as $T\to 0$. In this way an extremal dynamics is generated, namely
the quenched version of the DBM \cite{qdbm,qdbm1}.

(iii) By setting $\eta=0$, the dependence on the physical field is
canceled and we have a purely geometrical process. In fact, in 
that limit we obtain 
the IP extremal dynamics. For $T>0$ a characteristic
 length is generated in the dynamics as we will show later by
performing numerical simulations, 
and one observes asymptotically compact clusters, i.e. EDEN
\cite{EDEN} growth.

It is worth to stress that the previous limiting cases are quite 
general. For example one could also consider continuous
degrees of freedom $\sigma_i$ taking values on a 
range $[-x_i,x_i]$. The specific
equations change but the limiting form of $P_i$ given
by Eq. (\ref{pi}) for $T\gg 1$ and $T\ll 1$,  
remains unchanged.

\begin{figure}
\centerline{\psfig{figure=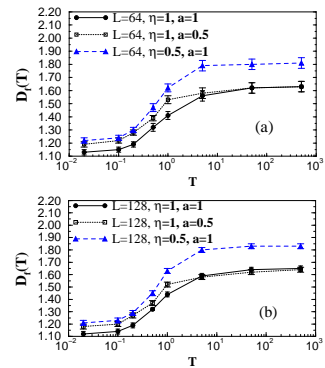,height=5cm,angle=0}}
\caption{ a) fractal dimension $D_f(T)$ of GDBM vs.
 the parameter $T$ (dimensionless)
 for system size $L=64$ (a) and $L=128$
 (b) with different values of $\eta,a$.
 Note the irrelevance of $a$ (disorder) for high temperatures.}
\label{fig2}
\end{figure}

In order to study the properties of our model in more detail 
in generic cases,
 we have 
performed numerical simulations.
Each bond of the lattice is assigned a growth probability 
as in Eq. \ref{pi} and the dynamics starts from the lowest 
electrode, which represents the seed of the growth process. 
As soon as the fracture pattern reaches the
 upper limit of the lattice, the dynamics is stopped, and a new 
realization starts.
We performed a set of about $50$ realizations of the dynamics 
of size $L \times 4L$ with $L\!=\!64, 128$, for a wide range of 
values of the temperature $T$, and different values of $\eta$ and $a$.
 For each set 
of realizations we compute by using
 a box-counting method \cite{bc} the fractal dimension, $D_f$, 
of the resulting clusters. Data are collected only in
 the central $L \times 2L$ part of the lattice, which is sufficiently far
from the lower transient regime and the upper interfacial region 
\cite{fst2}.
 In Fig. \ref{fig2}a-b we show a
plot of $D_f(T)$ vs. $T$ for $L\!=\!64$ and $L\!=\!128$ respectively 
and different values of $a$ and $\eta$, 
while in Tab.\ \ref{tab1}
we give some numerical values of $D_f(T)$ (for $\eta\!=\!1,a\!=\!1$). 
caca

\begin{table}
\begin{tabular}{|c|c|c|c|c|c|} \hline
$T$ &  $0.1$ & $0.5$ & $1.0$
& $5.0$ &  $500.0$ \\ \hline
$D_f(T) [L=64]$   &$1.15(2)$ 
&$1.32(3)$  &$1.41(3)$ & $1.56(3)$ 
& $1.63(4)$\\ \hline
$D_f(T) [L=128]$    &$1.14(2)$ 
&$1.32(1)$  &$1.44(2)$ & $1.59(2)$ 
& $1.65(2)$\\ \hline
\end{tabular}
\caption{Fractal dimension of aggregates of size $L=64,128$
for different values of $T$ ($a=1,\eta=1$).}
\label{tab1}
\hfill
\end{table}

It can be verified that, as predicted,
in the limits $T \to 0$ and $T \to \infty$ the model 
approaches the QDBM and DBM known values of $D_f$, $1.15$ and $1.66$
respectively 
\cite{dbm,qdbm,qdbmlet,fst2}.  {\it The transition from the QDBM-like 
behavior to the DBM-like one is smooth and continuous 
(Fig.\ \ref{fig2}), with continuously varying $D_f(T)$}.
We have checked that the critical properties of the model become 
less and less sensible to the value of $a$ as the 
temperature $T$ is raised (see Fig. 2), in agreement 
with our previous
arguments. In our picture of the breakdown phenomena, 
this means that for very high temperatures, {\it the thermal disorder 
is more relevant that the quenched one}. 

\begin{figure}
\centerline{\psfig{figure=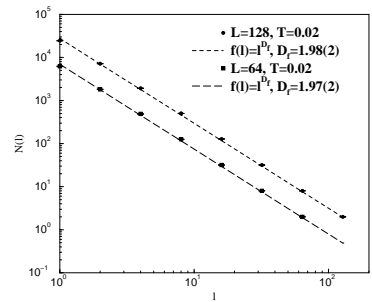,height=4cm,angle=0}}
\caption{
 Results of the box-counting
 analysis for a very small temperature $T=0.02$ and $\eta=0$.
 The values $D_f=1.97(2)$ for $L=64$ and 
$D_f= 1.98(2)$ for $L=128$ we get correspond to asymptotically 
compact clusters and, are very far from the IP fractal 
dimension $D_f \sim 1.89$.}
\label{fig3}
\end{figure}

When the Laplacian field is removed ($\eta=0$) the properties 
of the model change dramatically. We  observe a transition 
from an IP-like behavior to an EDEN like 
behavior. For a value of $T=0.02$ (the smallest value that we are
able to implement in
simulations),  we get already compact growth, with $D_f=1.98\pm0.02$ 
for size $L=128$; observe that the slight deviation from $D_f=2$
 being
a finite size effect)
(see Fig.\ \ref{fig3}) 
This result is not affected by changes of the disorder strength.
In fact, it can be easily shown that the extremal dynamics 
we recover in the $T=0$ limit, is independent on the value 
of $a$ if there is no Laplacian field \cite{qdbmlet,rtslung}.

\begin{figure}
\centerline{\psfig{figure=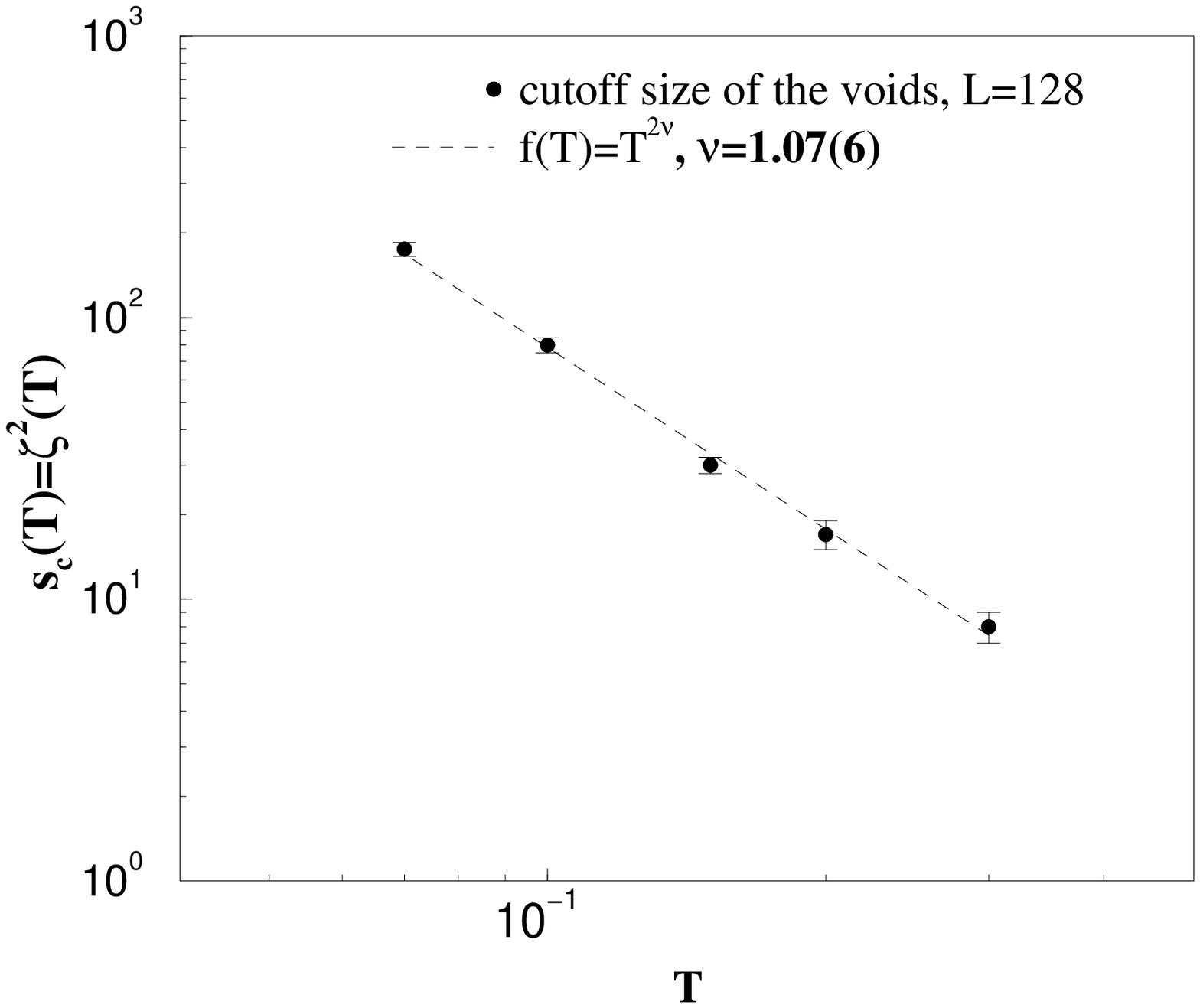,height=4.0cm}}
\centerline{\psfig{figure=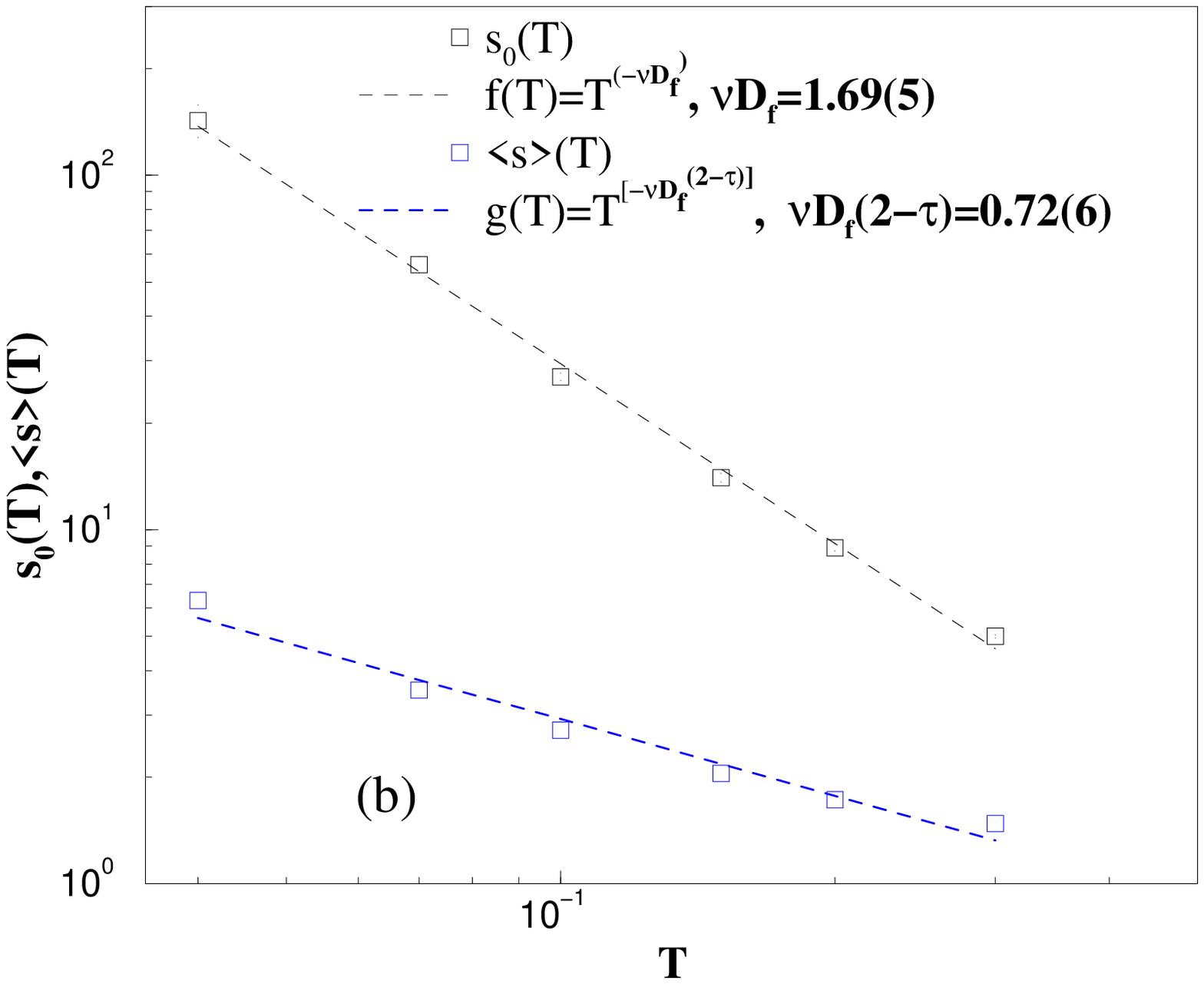,height=4.0cm}}
\caption{ (a): $\log_{10}$-$\log_{10}$
 plot of $s_c(T)=\xi^2(T)$ versus $T$ for 
the GDBM model without field.
 We find $\xi(T)\sim T^{-\nu}$
  with $\nu=1.07(6)$.
 We interpret this as an evidence of a 
 second order phase transition at $T_c=0$.
 (b): Scaling of $s_0$ and $\langle{s}\rangle$ vs. $T$.}
\label{fig4}
\end{figure}

To analyze in more detail the crossover from IP to EDEN model we 
studied the correlation properties of the clusters in the 
following way. A fractal structure
 is characterized by the presence of voids of all sizes with
no characteristic length; that
is precisely the
situation for $T=0$ in our model. However, for $T>0$ 
voids are still present in the clusters but they have 
a characteristic, temperature dependent, size $s_c(T)$.
 Since voids are compact objects, the square root of $s_c(T)$
represents the correlation length $\xi(T)$ of our model.
 We have studied the void distribution of our model for 
different small $T$ and system size $L=128$. By collapse 
plot techniques we have obtained the behavior
 of $s_c(T)=\xi^2(T)$
 shown in Fig.\ \ref{fig4}, and we have found a power law divergence of 
the correlation length as we approach $T=0$,
 with an associated {\em anomalous} 
exponent $\nu = 1.07(6)$. 
This power law divergence of the correlation length can
be interpreted as the hallmark of 
a second order phase transition along the temperature 
axis, with $T_c=0$. This is in line with recent results 
found for other self-organized  models \cite{ddrg}.
 To test this hypothesis we 
studied the avalanche distribution $D(s)$  for different $T$'s 
(for a definition of avalanches see for example 
\cite{rtslung}).
 This distribution is fitted by the scaling function:
$D(s)=s^{-\tau}f(s/s_0)$
where $s$ is the avalanche size, $\tau$ is the avalanche exponent, $f(x)$
 goes to a constant for $x\to0$ and falls exponentially for $x\to \infty$, and
$s_0$ is the cutoff size of the avalanches away from the critical point. 
By assuming 
$s_0 \sim T^{-\frac{1}{\sigma}} \sim \xi^{D_f}$, and 
$\langle s \rangle \sim T^{-\gamma}$,
where $\langle{s}\rangle=\sum_1^{\infty} s D(s)$,
 we obtain the usual scaling
relations among avalanche exponents 
$\frac{1}{\sigma}=\nu D_f \, , \, \gamma=
 \frac{(2-\tau)}{\sigma}$.
Our simulations give (Fig.\ref{fig4}): 
$1/\sigma=1.69(5)$,
 $\gamma=0.72(5)$,
 $\tau=1.60(2)$, $D_f=1.87(3)$. These values together with the above 
estimation of $\nu$ are in good agreement with the 
scaling relations among exponents, and 
support the interpretation of the transition
 at $T=0$ from IP to EDEN model as a second order 
dynamical phase transition
(a more detailed report on numerical as well as
analytical results will be given
 in a forthcoming paper.

In summary, we have presented a general model which include
DBM, QDBM, IP and EDEN models as limiting cases, and permits to 
interpolate among them by describing the 
the interplay between quenched disorder,
 thermal disorder and external stress fields.
 This represents a step towards an 
unification of the models actually used to 
describe fractal growth phenomena, under a common picture. 
We report a continuous change in the fractal dimension
as temperature is risen interpolating from QDBM to  DBM, 
and a novel phase transition from the extremal (self-organized)
IP fixed point to EDEN type of growth.
This  can trace a connection between fractal 
growth dynamics and ordinary critical phenomena.

 {\bf ACKNOLEDGMENTS} This work has been partially supported by
TMR Network, contract number FMRXCT980183, and 
TMR grant ERBFMBICT960925 to M.A.M.


\begin{thebibliography}{99}

\bibitem{dla}
T. A. Witten and L. M. Sander, {\em Phys. Rev. Lett.} {\bf 47}, 
1400 (1981). 

\bibitem{dbm}
L. Niemeyer, L. Pietronero and H. J. Wiessmann, 
{\em Phys. Rev. Lett.} {\bf 52},1033 (1984).

\bibitem{wilk}
D. Wilkinson and J. F. Willemsen, {\em J. Phys.} 
{\bf A 16}, 3365 (1983).

\bibitem{rtslung} R. Cafiero, A. Gabrielli, M. Marsili and 
L. Pietronero, {\it Phys. Rev.} 
{\bf  E 54}, 1406 (1996);
M. Marsili and M. Vendruscolo, Europhys. Lett. {\bf 37}, 505 (1997).


\bibitem{fst2}
A. Erzan, L. Pietronero and A. Vespignani, {\em Rev. Mod. Phys.}
{\bf 67}, 3 (1995).
                           

\bibitem{mem} M. Marsili, G. Caldarelli, and M. Vendruscolo, 
{\em Phys. Rev.} {\bf E 53}, R1 (1996).

\bibitem{qdbmlet}
R. Cafiero, A. Gabrielli, M. Marsili, L. Pietronero and L. Torosantucci,
 {\em Phys. Rev. Lett.} {\bf 79}, 1503 (1997).


\bibitem{EDEN}
M.Eden, {\em 4th Berkeley Symp. on Mathematical Stat. and Prob.},
pag.223 (1961).

 
\bibitem{ddrg}
 G.Grinstein, in {\em Scale Invariance, Interfaces and
 Nonequilibrium Dynamics}, Vol.344 of NATO ASI, series B, ed.s
A.McKane {\em et al.} (Plenum, New York, 1995).
A. Vespignani, R. Dickman, M. A. Mu\~noz, and S. Zapperi, 
Pre-print 1998.
 

\bibitem{qdbm} F. Family, Y. C. Zhang and T. Vicsek, 
{\em J. Phys.} {\bf A 19}, L733 (1986).

\bibitem{qdbm1} L. De Arcangelis, A. Hansen, 
H. J. Herrmann and S. Roux, {\em Phys. 
Rev.} {\bf B 40}, 877 (1989).

\bibitem{bc} K. J. Falconer, {\em Fractal Geometry: Mathematical
Foundation and Application} J. Willey and sons, New York, 1990.


\end{thebibliography}
 \end{document}